\documentclass[lettersize,journal]{IEEEtran}
\usepackage{amsmath,amsfonts}
\usepackage{algorithmic}
\usepackage{algorithm}
\usepackage{array}
\usepackage[caption=false,font=normalsize,labelfont=sf,textfont=sf]{subfig}
\usepackage{booktabs}
\usepackage{textcomp}
\usepackage{stfloats}
\usepackage{multirow}
\usepackage{enumitem}
\usepackage{amsmath,amsfonts,amssymb}
\usepackage{url}
\usepackage{hyperref}
\usepackage{verbatim}
\usepackage{graphicx}
\usepackage{cite}
\usepackage{xcolor}
\usepackage{algorithm}
\usepackage{algorithmic}
\begin{document}

\title{Unleashing Expert Opinion from Social Media for Stock Prediction}

\author{
    Wanyun Zhou\IEEEauthorrefmark{1}, 
    Saizhuo Wang\IEEEauthorrefmark{2,3}, 
    Xiang Li\IEEEauthorrefmark{1}, 
    Yiyan Qi\IEEEauthorrefmark{3}, 
    Jian Guo\IEEEauthorrefmark{3},
    Xiaowen Chu\IEEEauthorrefmark{1},~\IEEEmembership{Fellow,~IEEE}~\thanks{Corresponding authors: Jian Guo, Xiaowen Chu. Email: guojian@idea.edu.cn, xwchu@hkust-gz.edu.cn}\\
    \IEEEauthorblockA{\IEEEauthorrefmark{1}
        \textit{The Hong Kong University of Science and Technology (Guangzhou), Guangzhou, China}}\\
    \IEEEauthorblockA{\IEEEauthorrefmark{2}
        \textit{The Hong Kong University of Science and Technology, Hong Kong SAR}}\\
    \IEEEauthorblockA{\IEEEauthorrefmark{3}
        \textit{IDEA Research, Shenzhen, China}
    }
}

\maketitle

\begin{abstract}
While stock prediction task traditionally relies on volume-price and fundamental data to predict the return ratio or price movement trend, sentiment factors derived from social media platforms such as StockTwits offer a complementary and useful source of real-time market information. However, we find that most social media posts, along with the public sentiment they reflect, provide limited value for trading predictions due to their noisy nature. To tackle this, we propose a novel dynamic expert tracing algorithm that filters out non-informative posts and identifies both true and inverse experts whose consistent predictions can serve as valuable trading signals. Our approach achieves significant improvements over existing expert identification methods in stock trend prediction. However, when using binary expert predictions to predict the return ratio, similar to all other expert identification methods, our approach faces a common challenge of signal sparsity with expert signals cover only about 4\% of all stock-day combinations in our dataset. 
To address this challenge, we propose a dual graph attention neural network that effectively propagates expert signals across related stocks, enabling accurate prediction of return ratios and significantly increasing signal coverage. Empirical results show that our propagated expert-based signals not only exhibit strong predictive power independently but also work synergistically with traditional financial features. These combined signals significantly outperform representative baseline models in all quant-related metrics including predictive accuracy, return metrics, and correlation metrics, resulting in more robust investment strategies. We hope this work inspires further research into leveraging social media data for enhancing quantitative investment strategies. The code can be seen in https://github.com/wanyunzh/DualGAT.

\end{abstract}


\section{Introduction}

\textcolor{black}{Quantitative investment research has traditionally relied on handcrafted features from volume-price data and fundamental indicators for stock prediction. Deep learning has significantly advanced this field by moving beyond manual feature engineering to learn complex patterns directly from raw data. Early sequence models such as LSTMs \cite{lstm1,lstm2} and GRUs \cite{agree_tkde,gru1} were designed to capture temporal dependencies by using the financial time series, while subsequent developments introduced Transformer-based architectures \cite{gezici_deep_2024,tuncer_asset_2022} for enhanced sequence modeling and incorporated graph structures using GNNs to exploit inter-stock relations \cite{feng2019temporal,fingat_tkde,tkde_Spillover}. In recent years, there has been a growing focus on utilizing sentiment features derived from social media as a complementary information source for stock prediction \cite{acl2018stock,Deep_attentive_learning,acl2023causality,tkdd_ynamic}.}
Social media platforms for investment discussions have become increasingly influential in financial markets \cite{audrino2020impact,wang2023methods}. These platforms reflect real-time investor sentiment and demonstrate significant market-moving potential, exemplified by the GameStop short squeeze \cite{noauthor_gamestop_2024} that originated on social media and caused unprecedented volatility. Recognizing this potential, professional trading firms (e.g., WorldQuant) actively incorporate social media data into trading strategies for predictive signals \cite{worldquant_data}.
Among all investment-related social media platforms, StockTwits \cite{stocktwits} stands out as a specialized platform for investors to share stock insights. Each stock has its dedicated discussion thread where users post related messages. Additionally, StockTwits offers a unique advantage: users can explicitly label their posts as ``Bullish" or ``Bearish," providing clear sentiment indicators for specific stocks. This self-labeling mechanism eliminates algorithmic sentiment analysis uncertainty and offers more reliable sentiment signals.

\begin{figure}
    \centering
    \includegraphics[width=0.95\linewidth]{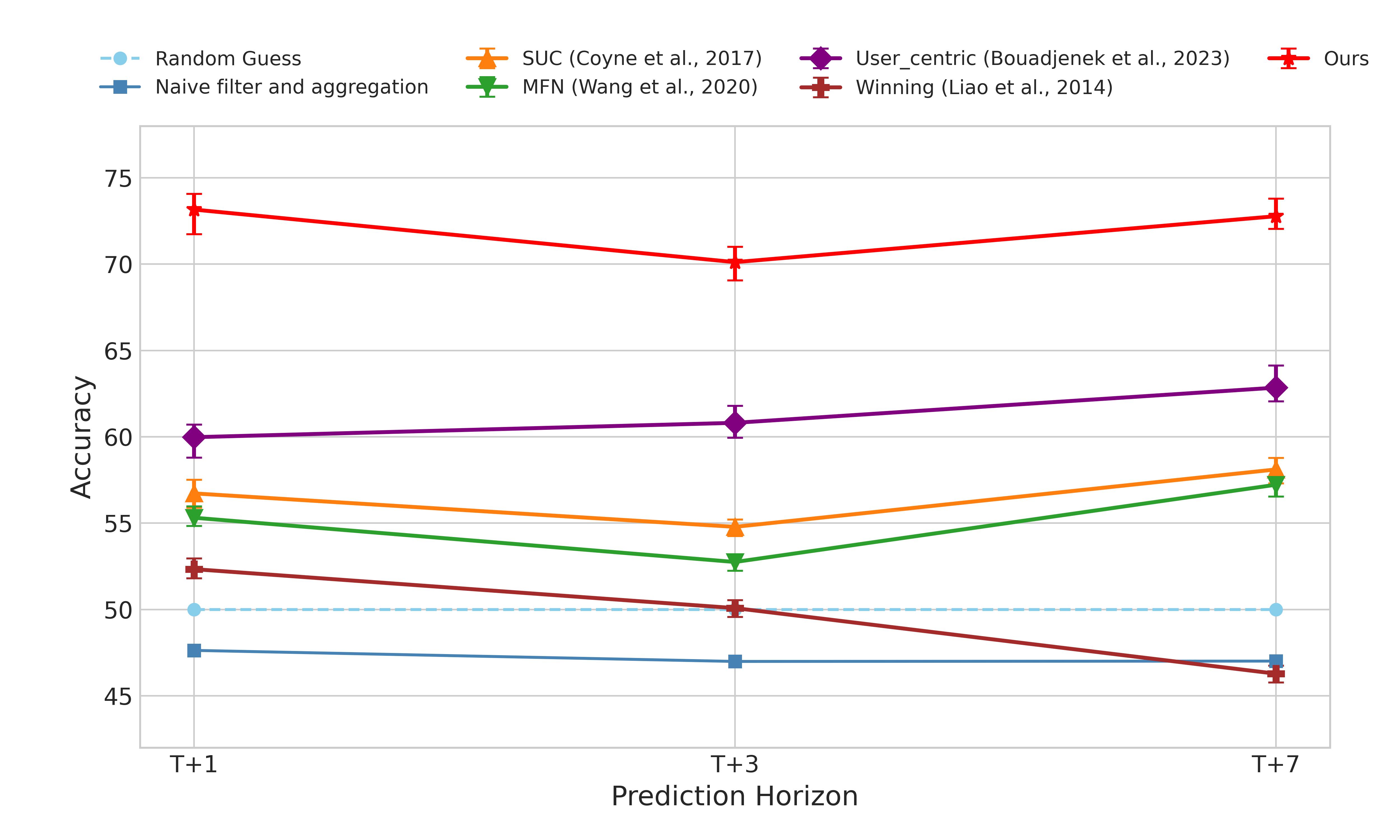}
    \caption{Comparison of stock price movement prediction accuracy (\%) achieved by experts identified through different expert identification methods versus naive sentiment aggregation methods. Accuracy is measured by how well subsequent sentiment predictions from identified experts align with actual stock price movements.}
    \label{fig:pre_experiment1}
\end{figure}

Although StockTwits provides valuable self-labeled sentiment data and contains a vast amount of discussions, extracting useful trading signals for either stock movement trend or return ratio prediction from the raw social media data remains challenging due to its substantial noise. To demonstrate that most social media posts are noisy and non-informative for trading predictions, we collected all StockTwits posts from 2016 to 2023 and examined the relationship between public sentiment and subsequent stock price movements. We applied a native filter and aggregation method that identifies the dominant sentiment for each stock based on a threshold of posts and the intensity of the sentiment. Specifically, if the number of posts for a given stock exceeded 30 on a given day, and one sentiment (either bullish or bearish) dominated by more than 85\%, we used that sentiment to predict the stock's trend for the following day. As shown by the solid blue line with square markers in Figure \ref{fig:pre_experiment1}, the accuracy of dominant sentiment for predicting the stock's movement trend (bullish for rise and bearish for fall) on the next day (T+1) was 47.63\%, 3 days later (T+3) was 46.99\%, and 7 days later (T+7) was 47.01\%. These results suggest that using aggregated sentiment for stock prediction yields accuracy that is not even better than random guessing, demonstrating the limited predictive value of social media sentiment. Such underperformance indicates that the majority of social media posts contain significant noise rather than actionable trading information.

In fact, social media data is inherently noisy when used as trading signals \cite{sentiment_analysis_for,Forecating_stock_prices,aaai_expert_based,xie2020signal,schnaubelt2020separating}. This stems from the open nature of these platforms, where anyone can post opinions that often lack predictive power. Retail investors' discussions, for example, have limited impact on stock movements, and many of their posts are speculative, driven by personal biases, or based on rumors \cite{rumor_volatility}. These non-professional posts often lack investment insights, making them unreliable as trading signals \cite{harvard_retail}. Additionally, there is potential for insider traders to manipulate social media to influence investor sentiment and, in turn, affect market trends \cite{tan2023market}. Moreover, the prevalence of emotional and impulsive posts during market volatility further contributes to this noise, as users tend to react to price movements rather than provide predictive insights \cite{ge2020beyond}. As a result, truly informative posts from expert users are rare, making it crucial to accurately identify these experts amidst the noise \cite{aaai_expert_based}.

Despite efforts in existing research to identify expert signals \cite{sentiment_analysis_for,A_user_centric,aaai_expert_based,Forecating_stock_prices,liao2014winning}, current methods struggle to reliably distinguish true experts from users who simply appear to be experts due to random correlations with market movements. These methods often overlook the fact that true experts should be able to make precise predictions across different market regimes. Their expert identification approaches tend to favor users who post frequently and whose posts happen to be correct during specific market trends, leading to biased results. Furthermore, these methods do not address the widespread presence of bots and spam accounts in social media posts, further undermining their effectiveness in identifying genuine expert signals.

To address these limitations in expert identification, we propose a novel dynamic expert tracing algorithm tailored to stock data. Our algorithm incorporates multiple key principles to ensure robust expert identification. First, it filters out noise from bots and spammers to address the data quality issue. Second, it evaluates both long-term and dynamic prediction accuracy across various market regimes to identify experts. Third, it ensures that experts focus only on a manageable number of stocks by filtering out users with numerous posts on different stocks in a single day. Furthermore, we identify both true experts and `\textbf{inverse experts}'—users who consistently make incorrect predictions. The identification of inverse experts are valuable trading signals for two main reasons. First, research in behavioral economics \cite{Hilton2001ThePO,lekovic2020cognitive} shows that several cognitive biases can lead to systematic misjudgments in market predictions, a characteristic frequently observed in inverse experts. This is particularly relevant since most traders in the market consistently lose money due to these cognitive biases, effectively making them inverse experts whose predictions can serve as reliable contrary indicators. 
\textcolor{black}{Second, inverse experts may emerge in contexts where market participants attempt to influence stock prices, particularly when certain institutions or large investors use social media to create misleading signals.}
For example, when they want to sell, they use controlled media accounts to post bullish views to encourage retail investors to buy. This allows them to offload their shares at higher prices while they sell in large volumes, causing the price to drop (see \cite{sec2022} for cases). Conversely, when these institutions want to buy, they post bearish views to prompt retail investors into selling. This enables them to accumulate shares at lower prices before they drive the price up. These manipulated media accounts, though deceptive in intent, can act as reliable inverse experts that provide valuable trading insights. 

For true experts, we align our trading actions with their sentiment predictions (taking long positions for bullish signals and short positions for bearish ones), while for inverse experts, we take the opposite positions. To evaluate our expert identification method, we measure predictive accuracy at different horizons: the next day (T+1), 3 days later (T+3), and 7 days later (T+7). 
\textcolor{black}{
Here, a prediction is considered correct if the direction implied by the expert's sentiment (bullish = expected rise, bearish = expected fall, with inverse experts treated oppositely) matches the actual stock price movement over the corresponding horizon. Importantly, we identify experts separately for each time horizon, ensuring that experts for each horizon are selected based on their predictive performance within that specific timeframe and subsequently evaluated using the same corresponding horizon. When multiple experts post about the same stock on the same day, we randomly sample one expert's sentiment per stock-day combination to ensure unbiased evaluation. 
As demonstrated in Figure \ref{fig:pre_experiment1}, experts identified through our expert tracing system achieve substantially higher accuracy in predicting future stock movement trends across all three time horizons compared to existing expert identification methods.}

While predicting the stock future trend is important, the primary objective of quantitative investing is not just to identify trends but to maximize excess returns by selecting stocks with the highest potential for future profits. Thus, our model aims to predict the stock return ratio for the next trading day. To achieve this, we transform the expert predictions, which serve as binary trend indicators (rise or fall), into continuous return ratios that can be treated as expert signals. Once we use these expert signals to predict return ratios, we evaluate model performance using quantitative metrics including correlation metrics (e.g., information coefficient, rank information coefficient), return metrics (e.g., annualized return and Sharpe ratio), and predictive accuracy for comprehensive strategy evaluation.

\begin{figure*}
    \centering
    \includegraphics[width=0.95\textwidth]{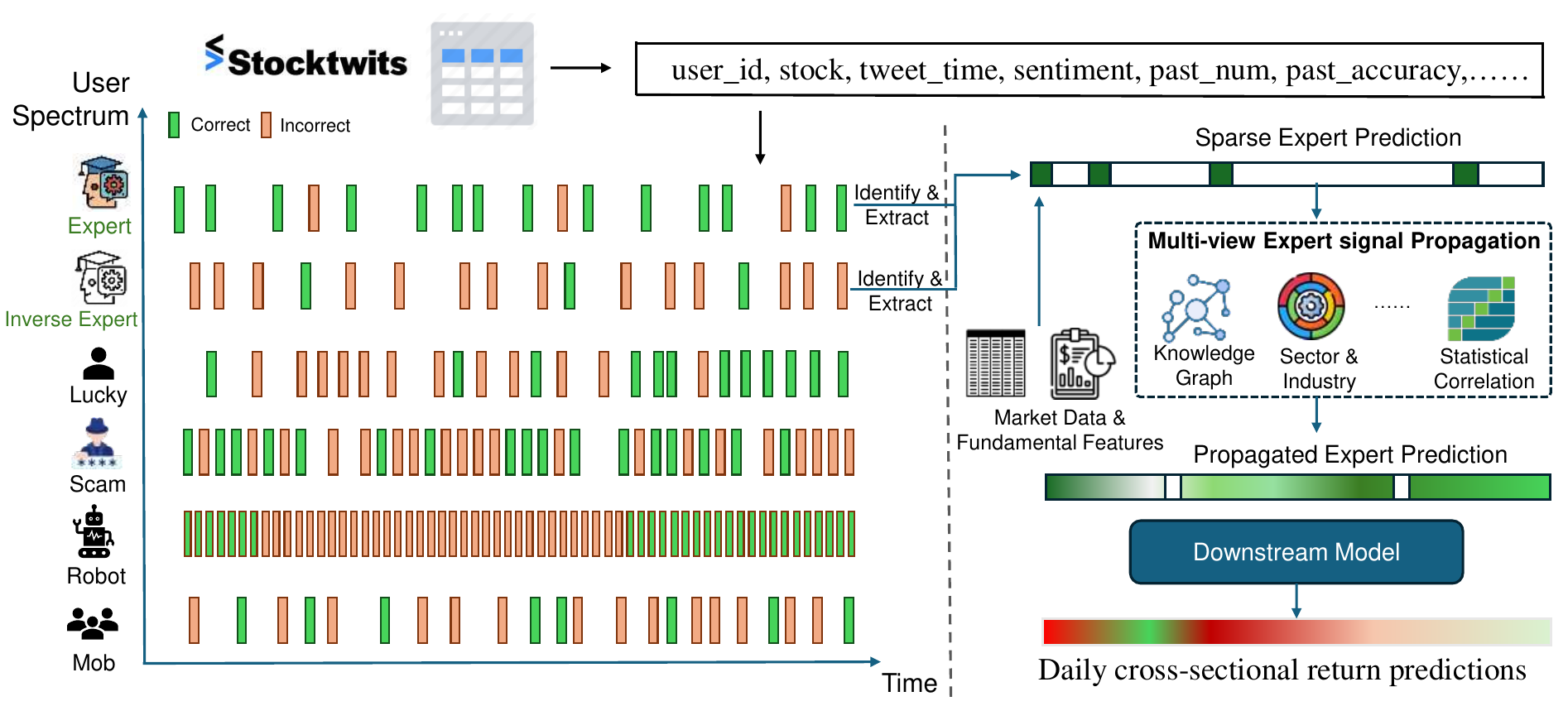}
    \caption{An overview of our proposed method. We first identify experts and inverse experts from social media based on historical prediction performance across different market regimes, as shown in the left part illustrating prediction patterns of different user types (experts/inverse experts, bots, spammers, lucky users, mob; see Sec.IV). Then we identify and extract these expert signals which are sparse in nature. To address the sparsity issue, we propose a dual graph attention network (DualGAT) that takes both sparse expert signals and market and fundamental features as input. DualGAT incorporates relational information among stocks from multiple aspects to propagate expert signals across related stocks, expanding expert signal coverage and finally outputs the daily cross-sectional return predictions for each stock.}
    \label{fig:workflow}
\end{figure*}

Though expert signals can provide strong predictions, challenges remain in their practical application. Like all methods that rely on expert predictions for trading signals, our strict expert tracing algorithm faces a common challenge of signal sparsity \cite{Forecating_stock_prices,A_user_centric,liao2014winning}. The resulting expert signals cover only about 4\% of all stock-day combinations in our dataset (i.e., most stocks have no expert predictions on most trading days), making it impractical for some quantitative investment scenarios, such as cross-sectional stock selection. To mitigate this sparsity issue, and considering the high dependency between asset returns in financial markets \cite{ane2003dependence,zhu2022forecasting,zhang2023graph}, we aim to leverage the powerful propagation capabilities of graph neural networks to enhance the flow of information \cite{zhang2023adaprop,xiao2021learning}. Therefore, we propose a message-passing mechanism based on dual graph attention networks, termed DualGAT, to propagate expert signals across related stocks. By using various sources of relational information on stocks to construct multiple graph structures, our approach effectively broadens the coverage of expert signals throughout the dataset. Additionally, when incorporating these expert signals with other traditional financial features (e.g., price-volume and fundamental data), the combined signals consistently demonstrate enhanced predictive power in forecasting stock returns. This synergistic effect suggests that our expert signals capture distinct stock characteristics that complement, rather than merely duplicate, the information of traditional financial features. 
Such complementarity is particularly valuable for quantitative strategies as it diversifies the sources of information, thereby enhancing the robustness of our investment framework. The pipeline of our method can be seen in Fig \ref{fig:workflow}.  In summary, our contributions are as follows:
\begin{itemize}
    \item We establish a comprehensive and up-to-date social media dataset, offering users easy access to several years of curated StockTwits data, eliminating the complexities of web scraping and making it more accessible for researchers and practitioners in the field.
    \item We address the issue of noisy social media data in stock prediction and propose a novel dynamic expert tracing algorithm that effectively identifies valuable trading signals by distinguishing both true experts and inverse experts from noise. Our approach substantially outperforms existing expert identification methods, with identified experts showing significantly higher accuracy in predicting stock movement trends.
    \item We address the common signal sparsity challenge in expert-based trading strategies through our DualGAT model, which greatly enhances signal coverage by propagating expert signals across related stocks.
    \item We demonstrate that our expert signals can work synergistically with traditional financial features, and these combined signals significantly outperform representative baseline models across all metrics, leading to more robust quantitative investment strategies.
\end{itemize}

\section{Related Work}
\subsection{Social Media-based Stock Prediction}
Stock prediction using social media data has gained significant attention in recent years due to its potential to capture market sentiment and crowd wisdom. Early research primarily focused on sentiment analysis of social media posts to predict market movements. For instance, Bollen et al. \cite{bollen2011twitter} analyzed Twitter mood to predict daily stock market changes, while Nguyen et al. \cite{nguyen2015sentiment} developed sentiment-based features from StockTwits messages for market prediction. These studies demonstrated that social media sentiment could serve as a valuable indicator of market trends. As deep learning advanced, especially in natural language processing, researchers began incorporating more sophisticated methods. DeepClue \cite{tkde2018deepclue} developed a hierarchical neural network to process company-related tweets from the social media and visualize their contribution to the prediction of stock movements.  Xu et al. \cite{acl2018stock} presented a market information encoder with a novel deep generative model to jointly encode tweets and price signals for stock prediction while Sawhney et al. \cite{Deep_attentive_learning} proposed MAN-SF that uses bilinear transformations to learn temporal interactions between tweet representations and historical prices. Zhang et al.\cite{tkdd_ynamic} integrated social media sentiment as one of the model inputs in their dynamic graph attention network.

However, as mentioned in the introduction, tweets from social media are inherently noisy. Moreover, our experiments (Figure \ref{fig:pre_experiment1}) show that public sentiment has limited predictive power for stock movements, which is also supported by Oliveira et al. \cite{On_the_predictability} and Coyne et al. \cite{Forecating_stock_prices}. While the aforementioned studies have made progress in leveraging social media data for stock prediction, they did not effectively filter the massive amount of tweets, potentially leading to models processing largely non-informative information. This highlights the importance of identifying experts within social media platforms whose predictions can provide informative and valuable trading signals.

\subsection{Expert Identification in Social Media}
Expert identification in social media has emerged as a critical research area, with various approaches developed to address this challenge. Early studies primarily employed authority-oriented methods or topic-oriented approaches \cite{zhao_expert_2015,zhao2016expert}. Authority-oriented methods maninly focus on users' reputation or social influence by constructing the user-to-user graph. For instance, Bouguessa et al. \cite{bouguessa_identifying_2008} employed an in-degree method to identify experts while Zhu et al. \cite{zhu_ranking_2014} ranked user authority in an extended category graph. Although effective, these approaches may not always match new questions with relevant expertise. In contrast, topic-oriented methods leverage latent topic modeling and linguistic analysis of users' posts to match experts based on question content. Guo et al. and Zhou et al.  \cite{guo2008tapping,zhou2012topic} introduced topic-based probabilistic models to question-answering activities and identify experts. Weng et al. \cite{weng2010twitterrank} proposed TwitterRank, which identifies influential users within specific domains by performing a topic-sensitive random walk that models the transition probability from one Twitter user to another. Additionally, temporal dynamics have been recognized as crucial factors in expertise identification, as demonstrated by Chi et al. \cite{chi2018expert}, who showed that dynamic LDA topic models outperform traditional static approaches in tracking evolving expertise indicators.

In the context of financial social media, expert identification primarily focuses on users' prediction accuracy for securities. Several studies \cite{aaai_expert_based,Forecating_stock_prices,sentiment_analysis_for,liao2014winning,A_user_centric} have attempted to identify financial experts based on their historical prediction accuracy over a specific time period or number of posts. However, these methods fail to filter out automated bots or spammers and suffer from the problem of insufficient past data criteria when identifying expert users (see Sec.IV Expert User Identification for a detailed analysis). These limitations are effectively addressed by our dynamic expert tracing algorithm.

\subsection{Graph-based Methods in Financial Applications}
Due to the phenomenon of stock momentum spillover, Graph Neural Networks (GNNs) and graph modeling have demonstrated remarkable success in financial applications by effectively modeling complex relationships between market entities \cite{tkde_Spillover,tkde_bitype,thgnn,samba}. Recent studies have leveraged GNN architectures to model both market relationships and temporal dynamics. Chen et al. \cite{chen2018incorporating} proposed a collaborative model that employs LSTM and GCN. Feng et al. \cite{feng2019temporal} introduced a temporal graph convolution layer to capture stock relations in a time-sensitive manner, enabling the evolution of relational weights among connected edges over time. Hsu et al. \cite{fingat_tkde} developed FinGAT, which employs graph attention networks to identify profitable stock combinations by learning latent interactions among stocks and sectors. Zhao et al. \cite{tkde_bitype} proposed a bi-typed market knowledge graph approach with dual attention networks to capture momentum spillover signals between stocks and related entities such as executives through heterogeneous GNNs. Wang et al. \cite{wang2022hatr} introduced HATR-I, a hierarchical adaptive temporal relational model that formulate different views of domain adjacency graphs into a unified multiplex network and use the multi-stage relational matching for stock prediction. Ying et al. \cite{ying2020time} captured temporal relationships using both sequential features and attributes from stock documents through a time-aware relational attention network. Li et al. \cite{li2024forecasting} transform the price series into a graph structure using chart similarity to forecast turning points in stock price.

Different from these works that primarily focus on modeling inter-stock market relationships, our GNN framework not only captures the intricate dependencies among stocks but also serves as an effective mechanism for propagating sparse expert signals across related stocks, thereby addressing the critical challenge of signal sparsity in expert-based trading strategies.

\section{Problem Formulation}
In this paper, we first aim to identify expert users and extract their trading signals from social media posts of the current trading day. Based on these expert signals, combined with stock features from previous trading days, we then predict the stock return ratio for the next trading day. Specifically, given a set of stocks $\mathcal{S}$, for each stock $u\in\mathcal{S}$ on trading day $t$, we collect both price-volume features (open,high,low,close,volume) and fundamental indicators to form a feature vector $x_{u,t} \in \mathbb{R}^F$. We focus on predicting the stock return ratio as it normalizes the price variation between different stocks. The return ratio at day $t+1$ is defined as:
\begin{equation}
{r}_{u,t+1} = (c_{u,t+1} - c_{u,t})/c_{u,t}
\end{equation}
where $c_{u,t}$ is the closing price of stock $u$ at day $t$. To predict the return ratio $\{r_{u,t+1}\}_{u \in \mathcal{S}}$
for day $t+1$, our model takes three types of inputs:

\begin{enumerate}
    \item Historical market data: For each stock $u \in \mathcal{S}$, we use price and fundamental features $X_{u,t} = [x_{u,t-L}, \ldots, x_{u,t-1}] \in \mathbb{R}^{F \times L}$
 from day $t-L$ to $t-1$ with a rolling window of size $L$. Unlike most stock prediction papers that use data up to timestep $t$, we use price features until $t-1$ to avoid look-ahead bias, as the exact price for day $t$ are only available after market close.
    \item Expert signals: For each stock $u\in\mathcal{S}$, we perform real-time collection of StockTwits posts during day $t$. The collection process ends 5 minutes before market close, from which we extract expert signals $e_{u,t}$ if available.
    \item Dynamic stock graph: To capture the relationships between stocks, we construct a dynamic graph $G_t = (V_t, E_t)$ based on price data up to day $t-1$, where nodes $V_t = \mathcal{S}$ represent the complete set of stocks and edges $E_t$ capture their relationships (see Sec. V.C Dual Graph Attention Network for details).
\end{enumerate}

\section{Expert User Identification}

Despite the presence of noise in social media data, there is still potential to harness it by developing more sophisticated models. One promising avenue is through building an expert tracing system, as proposed by several studies that attempted to identify expert users within social media platforms. Among them, the MFN model proposed by 
 Wang et al.\cite{aaai_expert_based}, the Smart-User-Classification model by Coyne et al. \cite{Forecating_stock_prices}, the Winners model by Liao et al. \cite{liao2014winning} and  ``consistently correct user identification" by Bouadjenek et al.\cite{A_user_centric} aimed to classify expert users from social media posts that carry subjective sentiments. While the former three models rely on external sentiment analysis models to label tweets, which might introduce errors, our dataset from StockTwits contains user-labeled sentiments (bullish or bearish), which eliminates the need for external sentiment inference and reduces potential errors. However, when applied to our comprehensive dataset spanning the past five years, these methods all failed to identify experts/inverse experts who could consistently make (in)accurate predictions during the test period. Users identified as experts in the training phase did not perform reliably as experts in the test phase.

Several factors contribute to the failure of these models, and they are key reasons why these approaches do not work effectively in our data. Additionally, we propose solutions to address these issues in our model:

\begin{enumerate}[leftmargin=*,labelsep=3pt]
  
    \item Filtering spammers: Neither of these models applied effective filtering to remove noise generated by automated bots or spammers. We discovered that social media platforms contain posts from bots or spammers, characterized by repetitive posts at fixed intervals (e.g., publishing content around every 24 seconds), all carrying sentiment labels. These spammers can post hundreds or even thousands of tweets with sentiment labels per day about the same stock. These excessive posts distort sentiment analysis and need to be partially filtered. To address this issue, for each user-stock pair, we only retain the post closest to market closing time on any given day.
    
    \item Sufficient Past Data Criteria: Existing methods use insufficient criteria for expert identification, either using a fixed number of $N$ past posts (e.g., $N$=10) \cite{liao2014winning} or considering prediction accuracy only over a short period of $K$ days (e.g., $K$=40 or 90) \cite{A_user_centric,aaai_expert_based}. However, these methods proved unreliable. In our dataset, users who achieved over 80\% accuracy in their past 20 posts had an average accuracy of only 54.0\%, and users who achieved 80\% accuracy in the past 40 days had an average accuracy of only 53.9\% during the testing phase. Further analysis revealed that most users only have high prediction accuracy during specific periods, after which their accuracy declines significantly, and vice versa. For example, users who posted for more than 20 days in 2022 with an accuracy above 80\% averaged about 50\% in 2023, while users who had accuracy below 20\% in 2022 saw their accuracy rise to nearly 50\% in 2023. To better understand the shortcomings of existing approaches and address these limitations in expert identification, we propose two key criteria:

    \begin{itemize}[leftmargin=*,labelsep=5pt]
        \item \textbf{Precise prediction across different market regimes}: Some users consistently post only bullish predictions, while others post only bearish predictions. Those posting only bullish predictions may appear expert-like during upward trends but become inverse experts during market corrections. Similarly, consistently bearish users may seem expert-like during downturns but become inverse experts during upward trends. These users have little analytical value as they lack real market insight. To address this, we not only consider the accuracy of the past $N$ tweets to ensure expert dynamism, but also evaluate their long-term predictions over two years to verify sustained accuracy across market cycles. \textcolor{black}{While existing expert identification methods often rely on short evaluation horizons (e.g., 40 or 90 days), this approach is vulnerable to market cycles. This 2-year evaluation window is grounded in the seminal work of Jegadeesh \& Titman \cite{jegadeesh1993returns}, which demonstrates that strategies based on high 3- to 12-month past returns may experience significant reversals in the following year. Furthermore, \cite{carhart1997persistence, fama2010luck} also demonstrate that longer evaluation horizons are essential, as users appearing expert-like during short-term momentum phases can become inverse experts when market regimes shift.}

        \item \textbf{Focused Expert Evaluation}: True market experts typically specialize in a manageable number of stocks to maintain deep analysis and insight. However, some users post multiple bullish or bearish sentiments for numerous stocks on the same day. If the market happens to move in their predicted direction, they will appear highly accurate and be incorrectly identified as experts in previously mentioned expert identification methods. Such widespread, unfocused posting behavior contradicts the nature of genuine expertise, which requires concentrated attention and thorough analysis. 
        \textcolor{black}{This principle is grounded in cognitive science and financial research. Cognitive Load Theory recognizes the inherent limits on human information processing capacity \cite{miller1956magical}. Deep financial analysis is cognitively demanding, and studies show that specialization correlates with superior forecast accuracy. For example, Clement \cite{clement1999analyst} found that analysts with less complex portfolios (covering fewer firms) produce more accurate earnings forecasts. To ensure we identify truly focused experts, we require that a user's past 20 posts must span at least five different trading days \footnote{we tested various parameters and found that five different days yielded the best performance considering both the coverage and accuracy}. This filters out users who employ
        a ``shotgun" approach by spreading predictions across many stocks without demonstrating sustained analysis.}
    \end{itemize}
\end{enumerate}

Furthermore, previous models focused solely on identifying experts who consistently made correct predictions, ignoring the possibility of identifying inverse experts. We have added inverse experts to our approach, where we identify these inverse experts and take the opposite action based on their opinions.

By utilizing this approach, we developed a dynamic expert tracing system. For each trading day, if there is an expert whose past predictions meet the aforementioned criteria, we treat that expert's prediction for the day as a representative trend signal, serving as the directional indicator for whether the stock will rise or fall.

\begin{algorithm}[t!]
\caption{Expert Tracing System for Date $d$}
\label{algorithm}
\begin{algorithmic}[1]
\REQUIRE current date $d$, minimum post requirement $N$, minimum days span requirement $K$,
long-term accuracy threshold $P_1$, recent accuracy threshold $P_2$, long-term range $T$,
tweet set $C$ with $C_d \subset C$ denotes tweets posted on day $d$, $C^i \subset C$ denotes all tweets by user $i$
\ENSURE expert and inverse expert extraction for date $d$

\STATE $C_d \leftarrow \text{filter\_daily\_latest}(C_d)$ \COMMENT{Keep only the latest post per user-stock pair on day d}
\STATE $I_d \leftarrow \text{all users of } C_d$

\FOR{each user $i \in I_d$}
    \STATE \COMMENT{\textit{Stage 1: Evaluate Recent Performance}}
    \STATE $C^{i}_{\text{recent}} \leftarrow \text{get\_recent\_N\_posts}(C, i, N)$ \COMMENT{$C^{i}_{\text{recent}}$ denotes the most recent N posts before day $d$ for user $i$}
    \STATE $days^{i}_{recent} \leftarrow \text{count\_unique\_days}(C^{i}_{recent})$
    \IF{$days^{i}_{recent} < K$}
        \STATE continue
    \ENDIF
    
    \STATE $T^{i}_{recent} \leftarrow 0$; $F^{i}_{recent} \leftarrow 0$ \COMMENT{Recent accuracy counters}
    \FOR{each $c \in C^{i}_{recent}$}
        \STATE Check stock price change $\rho$ on the next trading day
        \IF{$(c \in \text{bullish} \land \rho \in \text{rise}) \lor (c \in \text{bearish} \land \rho \in \text{fall})$}
            \STATE $T^{i}_{recent} \leftarrow T^{i}_{recent} + 1$
        \ELSE
            \STATE $F^{i}_{recent} \leftarrow F^{i}_{recent} + 1$
        \ENDIF
    \ENDFOR
    \STATE $A^{i}_{recent} \leftarrow T^{i}_{recent}/(T^{i}_{recent} + F^{i}_{recent})$
    \STATE \COMMENT{\textit{Stage 2: Evaluate Long-Term Performance}}
    \STATE $C^{i}_{long} \leftarrow \text{get\_posts}(C_{d-T:d}, i)$ \COMMENT{Get all posts in past T years}
    \STATE $T^{i}_{long} \leftarrow 0$; $F^{i}_{long} \leftarrow 0$ \COMMENT{Long-term accuracy counters}
    \FOR{each $c \in C^{i}_{long}$}
        \STATE Check stock price change $\rho$ on the next trading day
        \IF{$(c \in \text{bullish} \land \rho \in \text{rise}) \lor (c \in \text{bearish} \land \rho \in \text{fall})$}
            \STATE $T^{i}_{long} \leftarrow T^{i}_{long} + 1$
        \ELSE
            \STATE $F^{i}_{long} \leftarrow F^{i}_{long} + 1$
        \ENDIF
    \ENDFOR
    \STATE $A^{i}_{long} \leftarrow T^{i}_{long}/(T^{i}_{long} + F_{i,long})$
    \STATE \COMMENT{\textit{Stage 3: Classify User}}
    \IF{$A^{i}_{recent} \geq P_2$ \AND $A^{i}_{long} \geq P_1$}
        \STATE Mark $i$ as expert and take long (short) positions for bullish (bearish) sentiment
    \ELSIF{$A^{i}_{recent} \leq (1-P_2)$ \AND $A^{i}_{long} \leq (1-P_1)$}
        \STATE Mark $i$ as inverse expert and take short (long) positions for bullish (bearish) sentiment
    \ENDIF
\ENDFOR

\end{algorithmic}
\end{algorithm}

{\color{black}
Our dynamic expert tracing system, summarized in Algorithm 1, identifies experts for any given trading day $d$. The system relies on a historical database of all user posts. The algorithm first filters to keep only the latest post per user-stock pair on day $d$. It then iterates through each user who posted that day. For each user $i$, the system performs a two-stage evaluation using their historical posting data.

First, in the recent performance evaluation stage, we assess the user's most recent $N$ posts ($N=20$). To ensure the user is active and focused, we require these posts to span at least $K$ days ($K=5$). We calculate $T^{i}_{recent}$ and $F^{i}_{recent}$, representing the number of correct and incorrect predictions in these recent posts. This gives us their recent accuracy, $A^{i}_{recent}$. Second, the long-term performance evaluation stage assesses the user's track record over the past $T=2$ years ($T=2$). We compute $T^{i}_{long}$ and $F^{i}_{long}$, the total counts of correct and incorrect predictions over this period, to determine their long-term accuracy $A^{i}_{long}$.
Finally, a user is identified as an expert if they meet two criteria: their recent accuracy $A^{i}_{recent}$ must exceed threshold $P_2$ (80\%), and their long-term accuracy $A^{i}_{long}$ must exceed threshold $P_1$ (65\%). Similarly, a user is classified as an inverse expert if $A^{i}_{recent}$ falls below $1-P_2$ (20\%) and $A^{i}_{long}$ is below $1-P_1$ (35\%). The rationale for these threshold choices and sensitivity analysis can be found in Supplementary Material S1.
}

The expert tracing system achieved an accuracy rate as high as 72.8\%. Furthermore, the inference stage has time complexity $O(\text{len}(C_d))$, where $\text{len}(C_d)$ is the number of posts on day $d$. Since our StockTwits database contains historical posting data and accuracy for each user, the inference phase simply traverses all posts once to update relevant user information. The operation scales linearly with total posts. The low complexity and high accuracy demonstrate our method's effectiveness in enhancing stock prediction by filtering noise and focusing on expert and inverse expert insights.

\section{Expert Opinion Propagation}
Despite integrating both expert and inverse expert signals, like other expert identification approaches \cite{aaai_expert_based,Forecating_stock_prices,A_user_centric}, our model still suffers from signal sparsity. These signals produce meaningful predictions for only few stocks on some trading days, covering merely 4\% of our dataset. Consequently, a strategy built directly on these sparse signals would suffer from high exposure and excessive concentration risk.
To address this limitation, we propose a graph-based method that transforms these limited expert predictions into practical signals and propagates them across related stocks, thereby extending coverage to inform all data points in the dataset.

\subsection{Trend Signals Transformation}
Though predicting whether a stock will rise or fall is important, the ultimate goal is not simply trend forecasting but maximizing excess returns through investment strategies. Feng et al. \cite{feng2019temporal} pointed out that traditional stock prediction methods focusing on classification (predicting direction) or regression (predicting price) are suboptimal because they fail to align with selecting stocks with the highest potential returns, which is the real objective for investors. Given that a more practical approach involves optimizing return ratios rather than focusing solely on trend direction, it becomes necessary to transform the expert prediction which is the binary trend indicators into continuous return ratios which can be treated as expert signals.

The transformation is as follows: 
for stocks predicted by experts to rise, we compute the average return ratio from days within the past 30 days when the stock showed an upward trend. Similarly, for a predicted downward trend, we use the average of the down days' return ratios within the same period. This 30-day window for the return ratio which acts as a monthly indicator is a common approach in financial modelling for stock prediction \cite{Forecasting_significant_stock,Stock_market_prediction,A_systematic_review} as it helps to smooth out daily volatility while providing a sufficient sample size to capture recent market trends. We use this calculated average as the expert signal. If a stock is not predicted by experts on any given day, then the expert signal for that stock is set to 0. \textcolor{black}{We compare this trend signals transformation approach against simply using binary signals (+1/-1), with experimental results demonstrating the superiority of our method (see Supplementary Material S2).}

 \subsection{Temporal Pre-training Model}
Now that we have obtained the sparse expert signals, the challenge remains that for over 96\% of stock-day pairs, we still lack any expert prediction for these pairs. To address this, we propose an initial step to train a temporal model, which will serve as a baseline for generating return ratio estimates. This step acts similarly to a pre-training process, where the temporal model captures patterns in stock movements without direct reliance on expert signals.

The model is trained using historical stock price data (open, high, low, close prices, and trading volume) and fundamental data. Unlike many studies following Feng et al.'s \cite{feng2019temporal, Deep_attentive_learning,wang2022hatr} method of normalizing the entire dataset across the whole time dimension (including test sets), which may lead to test set leakage, we normalize each feature daily across the stock cross-section.

\subsubsection{MS-LSTM}
Inspired by MS-RNN \cite{chung2022hierarchical}, we propose the pre-training model, termed MS-LSTM (Multi-Scale Long Short-Term Memory). Each scale is handled by an independent LSTM network, allowing the model to capture temporal dependencies at varying resolutions. The trained model generates a foundational return ratio estimate for each stock daily, which acts as a baseline for stocks that lack expert signals.


Given input sequence $\mathbf{X} \in \mathbb{R}^{N \times L \times d}$, where $N$ is the number of stocks, $L$ is sequence length, and $d$ is feature dimension, MS-LSTM processes the data as follows:

1. For each scale $s_i$ in $\mathcal{S} = [s_1, s_2, \dots, s_n]$, the model extracts a subsequence by sampling the input at intervals of $s_i$:
\begin{align}
    \text{Extract}(\mathbf{X}, s_i) &= [\mathbf{X}_{t=0}, \mathbf{X}_{t=s_i}, \mathbf{X}_{t=2s_i}, \dots, \mathbf{X}_{t=L-s_i}] \nonumber \\
     & \in \mathbb{R}^{N \times L/s_i \times d}
\end{align}

2. Each extracted subsequence is processed by its own LSTM network:
\begin{equation}
\mathbf{H}^{(i)} = \text{LSTM}_i(\text{Extract}(\mathbf{X}, s_i))
\end{equation}
where $\mathbf{H}^{(i)} \in \mathbb{R}^{N \times L/s_i \times h}$ represents the sequence of hidden states for scale $s_i$, and $h$ is the hidden dimension.

3. The final representation is obtained by first extracting the last hidden state from each scale ($\mathbf{H}^{(i)}_{t=L/s_i}$), averaging these states across all scales, applying layer normalization, and then transforming through an MLP to predict return ratios:
\begin{equation}
\mathbf{h}_{\text{mid}} = \text{LayerNorm}\left(\frac{1}{n}\sum_{i=1}^n \mathbf{H}^{(i)}_{t=L/s_i}\right) \in \mathbb{R}^{N \times h}
\end{equation}
\begin{equation}
\hat{\mathbf{r}} = \text{MLP}(\mathbf{h}_{\text{mid}}) \in \mathbb{R}^{N \times 1}
\end{equation}
where $\hat{\mathbf{r}}$ represents the predicted return ratios for all stocks.

This multi-scale architecture allows the model to capture both short-term and long-term dependencies in the input sequence and produces return ratio estimates for each stock. \textcolor{black}{The superiority of MS-LSTM can be seen in Supplementary Material S8.}

\begin{figure*}
    \centering
    \includegraphics[width=0.95\textwidth]{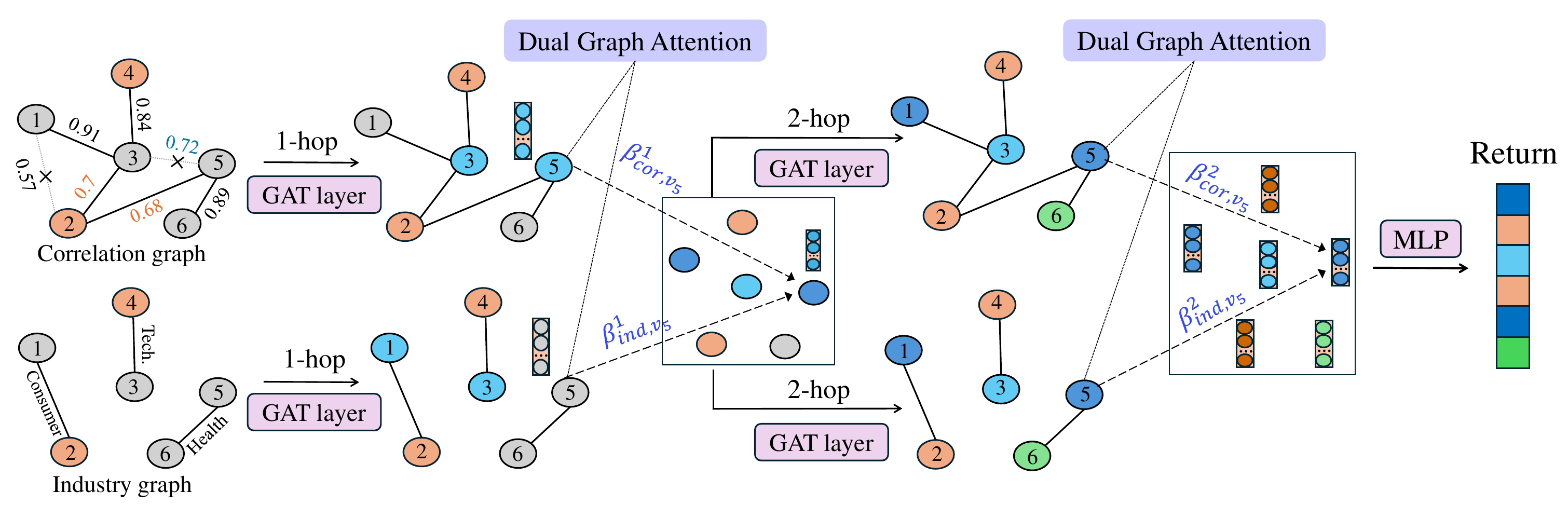}
    \caption{The workflow of DualGAT. On the far left, it illustrates the construction process of both the correlation and industry graphs, where the orange nodes represent stocks with initial expert signals. As seen in the figure, DualGAT facilitates the effective propagation of expert signals across all stocks.}
    \label{propagation_workflow}
\end{figure*}

\subsubsection{Loss Function}
During the pre-training process, we use the Information Coefficient (IC) loss function to train the network. As our task focuses on cross-sectional stock selection rather than timing individual stock movements, IC is particularly suitable because it measures the model's ability to rank stocks' relative performance, which is crucial for portfolio construction to outperform benchmark indices. Specifically, it calculates the correlation between the model's predicted output and the observed return ratio across the training set. The IC loss is computed as:
\begin{equation}
\text{IC} = \frac{\sum_{i=1}^{N} (\hat{r}_i - \overline{\hat{r}})(r_i - \overline{r})}{\sqrt{\sum_{i=1}^{N} (\hat{r}_i - \overline{\hat{r}})^2} \cdot \sqrt{\sum_{i=1}^{N} (r_i - \overline{r})^2}},
\end{equation}
where $\hat{r}_i$ and $r_i$ are the predicted and actual return for the $i$-th sample, and $\overline{\hat{r}}$ and $\overline{r}$ are the mean predicted and actual return, respectively. The IC loss function $\mathcal{L}_{IC}$ is defined as:
\begin{equation}
\mathcal{L}_{IC} = -\text{IC}.
\end{equation}
By maximizing the IC during training, the model is optimized to capture the correlation between its predictions and the actual observed returns.

\subsection{Dual Graph Attention Network}
For each stock on each day, the return ratio estimate generated from the pre-trained model is combined with two key features:

1. Expert availability: A binary feature indicating whether expert predictions exist for that day (1 if exist, 0 otherwise).

2. Expert signals: When an expert signal is available, the previously calculated expert signal value is used. If no expert prediction is made for a stock on a given day, it is set to 0.

These three features are combined to form a comprehensive feature set for each stock-day pair. Expert availability helps the model understand when to rely more heavily on the pre-trained output, while expert signals, when available, act as corrective signals to refine baseline predictions. Using this feature set, we propose using graph attention networks to propagate sparse expert signals and refine initial estimates from the temporal pre-training model.

For graph construction, we create two types of graphs to represent stock relationships: an industry graph and a dynamically updated correlation matrix graph. For the industry graph, nodes represent stocks, and edges are established between nodes if corresponding stocks belong to the same Global Industry Classification Standard (GICS) sector.
The correlation matrix graph is more dynamic, reflecting changes in stock relationships based on the previous 30 days of price movements. The following is the process for constructing the dynamic correlation matrix graph:

1. Data Preparation: Collect the past 30 days of stock price.

2. Correlation Calculation: Calculate the correlation coefficients for pairs of stocks and compute the correlation matrix based on the past 30-day window.

3. Graph Construction: Identify positively correlated stock pairs. If the correlation coefficient between two stocks exceeds the set threshold ($\theta_1 = 0.77$ as general threshold and $\theta_2 = 0.67$ for stocks with expert signals), an edge is created between this stock pair. For stocks with expert signals, we apply a lower correlation threshold, allowing these stocks to influence connected peers more strongly. \textcolor{black}{The rationale for these threshold choices and sensitivity analysis can be found in Supplementary Material S3.}
The leftmost diagram in Figure \ref{propagation_workflow} illustrates the construction process of both correlation and industry graphs. In the correlation graph example, an edge exists between nodes 2 and 5 as their correlation coefficient (0.68) exceeds $\theta_2$, with node 2 containing expert signals. However, despite having a higher correlation coefficient than nodes 2 and 5, nodes 3 and 5 are not connected because their correlation coefficient fails to reach $\theta_1$, as neither node contains expert signals.

The graph is updated daily to reflect the latest market dynamics. This adaptive approach ensures that our model can respond to changes in stock relationships and leverage new information as it becomes available. By incorporating both industry and correlation relationships, we significantly expand the coverage of expert signals beyond directly connected stocks, as signals can propagate through multiple paths in these complementary graphs. To effectively integrate and leverage these complex relationships, we implement a dual-graph attention-based graph neural network. Our GNN architecture, termed DualGAT, is constructed to process signals from both the industry and correlation graphs concurrently, enabling it to adaptively learn which graph provides more relevant information at each node for every given stock at each step. The whole framework of our DualGAT can be seen in Figure \ref{propagation_workflow} and it consists of several components structured to handle inputs from two distinct graphs:

The whole framework of our DualGAT can be seen in Figure \ref{propagation_workflow}. The model processes information from two distinct graphs through a two-hop mechanism with graph attention, followed by attentive feature fusion at each hop. The framework consists of the following key components:
\begin{enumerate}
    \item \textbf{Graph Attention Layers:} 
    For each graph (industry and correlation), we employ Graph Attention Networks (GAT) to learn node representations.
    \textcolor{black}{These GAT layers correspond to the pink blocks shown in Figure \ref{propagation_workflow}.}
    At each layer, a node $v$ aggregates information using attention mechanisms from its neighborhood $\mathcal{N}(v)$, where $\mathcal{N}(v)$ includes both the adjacent nodes and node $v$ itself (i.e., $v \in \mathcal{N}(v)$). The attention coefficients $\alpha_{vu}$ between node $v$ and each node $u \in \mathcal{N}(v)$ are computed as:
    \begin{equation}
        \alpha_{vu} = \frac{\exp(\text{LeakyReLU}(\mathbf{a}^\top [\mathbf{W}h_v \Vert \mathbf{W}h_u]))}{\sum_{k \in \mathcal{N}(v)} \exp(\text{LeakyReLU}(\mathbf{a}^\top [\mathbf{W}h_v \Vert \mathbf{W}h_k]))}
    \end{equation}
    where $h_v \in \mathbb{R}^{d_{in}}$ is the input feature vector for node $v$, $\mathbf{W} \in \mathbb{R}^{d_{hidden} \times d_{in}}$ is a learnable weight matrix, and $\mathbf{a} \in \mathbb{R}^{2d_{hidden}}$ is the attention vector. The node features are then transformed:
    \begin{equation}
        h'_v = \sigma\left(\sum_{u \in \mathcal{N}(v)} \alpha_{vu} \mathbf{W}h_u\right)
    \end{equation}
    where $h'_v \in \mathbb{R}^{d_{hidden}}$ is the output feature vector and $\sigma$ is a non-linear activation function (ReLU).

    \item \textbf{Dual-Graph Attentive Feature Fusion:}
After each graph attention layer, we perform attentive fusion of features from both graphs. Let $H_{ind} \in \mathbb{R}^{N \times d_{hidden}}$ and $H_{cor} \in \mathbb{R}^{N \times d_{hidden}}$ be the node features from industry and correlation graphs respectively, where $N$ represents the total number of stocks (i.e., the number of nodes in each graph).  The fusion process involves:

\begin{equation}
    \begin{aligned}
        \alpha_{ind} &= \sum_{j=1}^{d_{hidden}} q_{ind,j} H_{ind,j} \in \mathbb{R}^N \\
        \alpha_{cor} &= \sum_{j=1}^{d_{hidden}} q_{cor,j} H_{cor,j} \in \mathbb{R}^N
    \end{aligned}
\end{equation}

where $q_{ind}, q_{cor} \in \mathbb{R}^{d_{hidden}}$ are learnable attention vectors. \textcolor{black}{This computation corresponds to the connecting lines from the graph nodes to the purple "Dual Graph Attention" blocks in Figure \ref{propagation_workflow}.} The attention weights are normalized using softmax:

\begin{equation}
    [\beta_{ind}, \beta_{cor}] = \text{softmax}([\alpha_{ind}, \alpha_{cor}]) \in \mathbb{R}^{2 \times N}
\end{equation}
\textcolor{black}{These $\beta$ values correspond to the blue attention coefficients shown in Figure \ref{propagation_workflow}.}
The fused features are then computed as:

\begin{equation}
    \begin{aligned}
        H_{fused} = &(\beta_{ind}^\top \otimes \mathbf{1}_{d_{hidden}}) \odot H_{ind} \\ 
                    &+ (\beta_{cor}^\top \otimes \mathbf{1}_{d_{hidden}}) \odot H_{cor} \\
                    &\in \mathbb{R}^{N \times d_{hidden}}
    \end{aligned}
\end{equation}

where $\beta_{ind}^\top, \beta_{cor}^\top \in \mathbb{R}^{N \times 1}$ are the transposed attention weights, $\otimes$ denotes the Kronecker product with $\mathbf{1}_{d_{hidden}} \in \mathbb{R}^{d_{hidden}}$ (broadcasting to $\mathbb{R}^{N \times d_{hidden}}$), $\odot$ represents element-wise multiplication, and $H_{fused}$ represents the fused node features that will be used as input for the next graph attention layer or final prediction. \textcolor{black}{This fusion mechanism effectively combines information from both graph structures as visualized by the merging arrows in Figure \ref{propagation_workflow}.}

    \item \textbf{Two-Hop Architecture:}
    The model employs two consecutive hops of graph attention and feature fusion. The first hop transforms input features (with dimension being $d_{in}$ for each node) to hidden representations (with dimension being $d_{hidden}$ for each node), followed by feature fusion. The second hop processes these fused features to produce final representations (with dimension being $d_{out}$ for each node), which undergo another round of fusion to obtain $H_{fused}^{final} \in \mathbb{R}^{N \times d_{out}}$. Finally, a simple MLP layer transforms the fused features to scalar predictions:
    \begin{equation}
        y_v = \mathbf{W}_{mlp}h_{fused,v}^{final} + b
    \end{equation}
    where $h_{fused,v}^{final} \in \mathbb{R}^{d_{out}}$ represents the feature vector of node $v$ in $H_{fused}^{final}$, $\mathbf{W}_{mlp} \in \mathbb{R}^{1 \times d_{out}}$ and $b \in \mathbb{R}$ are the parameters of the final MLP layer.
\end{enumerate}

Our dual-graph attention mechanism enables effective information propagation through both industry relationships and correlation-based connections. The model adaptively learns the relative importance of each graph structure for different nodes during message-passing, leading to optimal feature fusion at each hop. Leveraging this adaptive fusion mechanism, the combination of dual-graph attention and expert signals allows the model to refine and correct initial predictions. By incorporating these sparse yet valuable signals, the synergy between temporal model output and expert information enhances the accuracy and robustness of information propagation across all stocks, ultimately improving predictive performance.

\section{Experiments}
\subsection{Experimental Setup}

\subsubsection{Dataset}
In our study, the expert signals come from StockTwits, a social media platform designed specifically for investors and traders to share insights. Users can post messages related to individual stocks, each of which has its own discussion thread for comments. These messages are often tagged with sentiments like ``bullish" or ``bearish," reflecting the user’s outlook on the stock. StockTwits merges social media engagement with stock market discussions, making it a valuable tool for capturing real-time investor sentiment.

We accessed data through the \href{https://api.stocktwits.com/api/2/streams/symbol/{stock}.json}{StockTwits API}, using the endpoint to fetch publicly available data related to specific stocks. During collection, we implemented an automatic pagination mechanism in our data retrieval script. This allowed us to capture complete historical data from the forum.

To optimize data retrieval efficiency, we employed a distributed system using proxy services, enabling multi-threaded, multi-node data collection. A single node collects one day's data in about 30 minutes. Using 30 nodes simultaneously can accelerate collection, retrieving daily posts within a minute. This speed increase ensured quick data gathering, enabling timely expert signal mining for subsequent model training.

The collected data contains valuable information, with each tweet representing a user's opinion about a specific stock. For each tweet, the data includes content, the referenced stock (industry, sector, symbol, and exchange), creation time, user ID, likes received, sentiment (bullish or bearish), and retweet information. Our data collection spanned from 2010 to 2023, and we established a database storing all collected data. This allows users to access StockTwits data directly without writing complex extraction scripts. Stock price data was sourced from \href{https://polygon.io}{Polygon.io}, and fundamental data from \href{https://eodhd.com}{EODHD}.

For the pre-training of our temporal model, we utilized the collected stock prices and fundamental data spanning from 2010 to 2018. To help enrich our expert signals collection and avoid isolated nodes in the industry graph, we augmented our dataset with the 10 stocks that possessed the most expert signals during the training phase. Our DualGAT model was trained and validated using data from 2019 to 2021 and conducted extensive testing of our DualGAT model on three distinct datasets: the Nasdaq 100 index, S\&P 500 index, and the updated StockNet dataset by \cite{www2024learning}, using data from 2022 to 2023 as the test sets.

\subsubsection{Baseline}
In this paper, we employ three different baselines to evaluate the effectiveness of our model:

\noindent \textbf{1. Expert signals baseline:} As illustrated in Figure 1, the expert selection methods proposed by Wang et al.\cite{aaai_expert_based} in the MFN model, the smart user classification model proposed by Coyne et al. \cite{Forecating_stock_prices}, the user centric method proposed by Bouadjenek et al. \cite{A_user_centric} and the Wining model proposed by Liao et al. \cite{liao2014winning} were selected as manually designed baseline models to assess the accuracy of the expert signals.

\noindent \textbf{2. Simple message propagation baseline:} To demonstrate the effectiveness of the message-passing capability in our designed DualGAT, we employ the simplest form of message passing without graph neural network architectures, denoted as $Wx$, where $W$ is the normalized adjacency matrix derived from either our industry or correlation matrix, and $x$ represents the stock's expert signal. Similar to DualGAT, this message-passing method also performs a two-hop propagation. This process helps to diffuse the expert signals to adjacent nodes, thereby leveraging the structural connections of the graph to enhance the predictive ability. This baseline is crucial for understanding the ability of the expert signals independently as a trading signal. At the same time, it also serves as the baseline for our DualGAT model.

\noindent \textbf{3. Temporal or spatial-temporal model comparison baseline:} To demonstrate the effectiveness of our expert-driven approach combined with DualGAT, this baseline employs other models known for temporal and spatial-temporal predictions in time series forecasting or stock prediction. We use these models to train and compare performance on the same training, validation and test sets.


\begin{table*}[htbp]
\centering
\caption{Performance of different propagation methods against the DualGAT model across various metrics on different datasets. ACC, IC, RIC, ICIR, AR are measured in percentage (\%).}
\label{tab:model_comparison}
\scalebox{0.83}{  
\begin{tabular}{c|cccccc|cccccc|cccccc}
\hline
& \multicolumn{6}{c|}{NASDAQ 100} & \multicolumn{6}{c|}{SP 500} & \multicolumn{6}{c}{Updated StockNet} \\
\cline{2-7} \cline{8-13} \cline{14-19}
Method & ACC & IC & RIC & ICIR & AR & SR & ACC & IC & RIC & ICIR & AR & SR & ACC & IC & RIC & ICIR & AR & SR \\
\hline
Correlation matrix propagation & 55.71 & 5.33 & 6.79 & 19.53 & 21.11 & 1.16 & 52.98 & 6.86 & 7.06 & 34.47 & 19.81 & 1.72 & 52.32 & 6.08 & 5.31 & 24.61 & 19.98 & 1.81 \\
Industry matrix propagation & 54.40 & 4.76 & 4.64 & 19.02 & 16.93 & 0.75 & 52.08 & 4.86 & 5.62 & 31.83 & 15.60 & 1.21 & 51.71 & 5.39 & 3.05 & 28.61 & 10.04 & 0.88 \\
Correlation+Industry matrix propagation & 56.62 & 6.25 & 6.42 & 26.62 & 21.18 & 1.45 & 52.91 & 7.13 & 6.30 & 41.99 & 21.78 & 2.09 & 52.82 & 6.12 & 3.32 & 24.15 & 17.81 & 1.16 \\
Our DualGAT model & \textbf{57.83} & \textbf{7.98} & \textbf{9.69} & \textbf{36.66} & \textbf{32.91} & \textbf{2.69} & \textbf{55.02} & \textbf{8.26} & \textbf{8.29} & \textbf{43.17} & \textbf{29.60} & \textbf{2.92} & \textbf{54.81} & \textbf{7.62} & \textbf{7.78} & \textbf{30.69} & \textbf{23.46} & \textbf{1.87} \\
\hline
\end{tabular}
}
\end{table*}

\subsubsection{Metrics}
For the evaluation of our model's performance, we select six key metrics: ACC (Accuracy), IC (Information Coefficient), RIC (Rank information coefficient), ICIR (Information Coefficient of Information Ratio), AR (Annualized Return), and SR (Sharpe Ratio).
\begin{itemize}
    \item \textbf{ACC}: This metric reflects the percentage of correct predictions our model makes regarding the upward or downward trend of stock prices.
    \item \textbf{IC and ICIR}: IC measures the correlation between predicted and actual values, quantifying the model's ability to make correct relative predictions rather than absolute stock price predictions. ICIR adjusts IC for signal volatility, providing a refined measure of predictive performance over time.
    \item \textbf{RIC}: RIC measures the correlation between the predicted rankings of stocks and their actual rankings. It quantifies the model's ability to correctly rank stocks relative to each other.
    \item \textbf{AR}: For returns, our strategy involves taking long positions in the top 10\% of stocks predicted to have the highest return ratio, and short positions in the 10\% predicted to decrease the most. \textcolor{black}{We apply a 4 basis points trading cost (0.04\%) to account for transaction costs and market impact.} This method is designed to maximize potential returns from market volatility.
    \item \textbf{SR}: The Sharpe Ratio measures risk-adjusted returns by dividing excess returns by return volatility, offering insights into the risk-adjusted performance of our investment strategy. 
\end{itemize}

 \textcolor{black}{The detailed mathematical definitions of these evaluation metrics can be found in Supplementary Material S4.} These metrics collectively provide a comprehensive framework for assessing the effectiveness and reliability of our predictive model in a dynamic financial environment.

\subsection{Main Results}
In this study, we have systematically evaluated our expert tracing system along with the DualGAT model against various models and baselines to determine its effectiveness and efficiency in stock price prediction.

For the mining of expert signals, Figure \ref{fig:pre_experiment1} illustrates the performance of several expert identification methods across three prediction horizons: T+1, T+3, and T+7. The model labeled ``Ours" significantly surpasses other models in accuracy at each time point. At T+1, ``Ours" shows an accuracy of 72.8\%, which is 13\% and 15\% higher than the second and third place models (User Centric \cite{A_user_centric} and SUC \cite{Forecating_stock_prices}). At T+3 and T+7, our method also demonstrates substantial leads over existing approaches. 
\textcolor{black}{
To further validate the robustness of our expert identification method, we applied it to different markets and social media platforms. We tested our approach on Xueqiu, which is the largest stock discussion platform in China with a stock pool comprising all stocks from the Shanghai and Shenzhen stock exchanges. 
Our method also substantially outperforms the leading alternative expert identification method from prior work, achieving approximately 10\% higher accuracy on this dataset. Detailed comparisons can be found in Supplementary Material S5. 
These results demonstrate that our expert identification method is highly effective across different social media sources and diverse market environments. Moreover, the consistent superior performance across various prediction horizons validates the robustness and reliability of our approach for practical investment applications. Additionally, we conducted an experiment to compare the performance of signals derived solely from identified true experts versus those from inverse experts, with the detailed analysis presented in Supplementary Material S6.}

To demonstrate the effectiveness of the message-passing capability in our designed model, we compare our DualGAT with the simple message passing method denoted as $Wx$ mentioned above. For the propagation matrix $W$,we utilized the correlation matrix ($W_{cor}$), the industry matrix ($W_{ind}$), and a combination of both. The result in Table \ref{tab:model_comparison} shows that DualGAT outperforms all message passing methods on all metrics, indicating DualGAT’s architectural advancements. 

Moreover, we demonstrate that DualGAT effectively addresses the sparsity of expert signals. Since the graph attention layer in our neural network corresponds to one hop of propagation, its signal propagation capability is comparable to simple message passing methods (similar to BFS, where edges in the two graphs can spread expert signals). Through empirical analysis using simple message passing methods, we show that a two-hop propagation setting achieves sufficient node coverage, reaching 89.1\% of all nodes. DualGAT enhances this propagation process by incorporating learnable attention coefficients, which dynamically adjust the importance of different nodes. Furthermore, attention mechanisms are applied between the industry and correlation graphs, enabling the model to adaptively weigh the contribution of each graph structure during the propagation process, thereby improving the prediction performance. \textcolor{black}{Our examination of the learned attention weights reveals substantial variation across different stocks between the two graphs, providing compelling evidence of the model's adaptive behavior. Detailed attention weight analysis can be found in Supplementary Material S7.}

When comparing with other temporal or spatial-temporal models used in time series forecasting or stock prediction shown in Table \ref{tab:performance_evaluation_models}, we find that our model significantly outperforms the traditional and more recent models in time series forecasting and stock prediction across all metrics on all datasets. \textcolor{black}{Notably, our signal propagation mechanism through our DualGAT architecture proves highly effective. Even when evaluated solely on the subset of stocks without direct expert signals (as shown in the `Ours w/o expert signal' row), it still significantly outperforms all baseline models, demonstrating successful signal propagation from the small fraction of stocks with expert signals to the broader stock universe.}

\textcolor{black}{To further validate robustness, we conducted out-of-sample testing across distinct market regimes spanning COVID periods (2019-2023). Our model consistently outperforms across these conditions, showing strong generalization capabilities. Detailed evaluation setup and cumulative return comparisons are in Supplementary Material S8.}

Since both our model and the baseline model use the same traditional financial features, the significant improvement of our model with the addition of expert signals demonstrates that our expert signals capture valuable information beyond what traditional financial features provide. Overall, these results affirm the superiority of our approach, especially in utilizing expert signals enhanced by graph-based learning.


\begin{table*}[htbp]
\centering
\caption{Performance comparison of various time series forcasting and stock prediction models on the test set. ACC, IC, RIC, ICIR, AR are measured in percentage (\%). / represents out of cuda memory.}
\label{tab:performance_evaluation_models}
\resizebox{\textwidth}{!}{
\begin{tabular}{c|cccccc|cccccc|cccccc}
\hline
& \multicolumn{6}{c|}{NASDAQ 100} & \multicolumn{6}{c|}{SP 500} & \multicolumn{6}{c}{Updated StockNet} \\
\cline{2-7} \cline{8-13} \cline{14-19}
Model & ACC & IC & RIC & ICIR & AR & SR & ACC & IC & RIC & ICIR & AR & SR & ACC & IC & RIC & ICIR & AR & SR \\
\hline
LSTM & 49.22 & 0.84 & 0.65 & 4.52 & 5.95 & 0.17 & 49.28 & 3.72 & 1.65 & 16.73 & 19.35 & 1.62 & 49.00 & 0.57 & 1.16 & 2.37 & 2.39 & -0.13 \\
Mixer \cite{tolstikhin2021mlp} & 49.81 & 1.00 & 1.27 & 5.88 & 7.38 & 0.38 & 50.72 & 3.70 & 2.02 & 19.64 & 19.76 & 1.79 & 51.06 & 1.44 & 1.64 & 5.14 & 5.86 & 0.15 \\
PatchTST \cite{nie2022time} & 49.58 & 1.90 & 1.04 & 7.22 & 12.38 & 1.24 & 50.36 & 1.76 & 1.89 & 15.19 & 9.49 & 0.91 & 49.85 & 2.49 & 2.51 & 11.35 & 17.36 & 1.47 \\
Crossformer \cite{zhang2023crossformer} & 49.34 & 0.76 & 0.56 & 3.14 & 9.36 & 0.43 & 49.28 & 2.07 & 0.38 & 10.04 & 7.11 & 0.40 & 50.95 & 0.01 & -0.05 & 0.04 & -2.11 & -0.52 \\
TimesNet \cite{wutimesnet} & 49.74 & 0.92 & 0.91 & 5.13 & 9.72 & 0.70 & 50.73 & 2.06 & 1.62 & 11.53 & 10.42 & 0.87 & 49.03 & 0.55 & 0.70 & 2.19 & 3.38 & -0.06 \\
MS-LSTM  & 49.40 & 0.89 & 0.88 & 4.81 & 6.94 & 0.54 & 49.29 & 2.84 & 2.96 & 18.45 & 17.39 & 1.29 & 49.06 & 0.85 & 1.46 & 3.52 & 9.57 & 0.74 \\
TimeMixer \cite{wangtimemixer} & 50.32 & 0.71 & 0.91 & 4.09 & 7.68 & 0.30 & 50.37 & 1.40 & 1.28 & 10.32 & 10.78 & 0.78 & 50.33 & 0.76 & 1.14 & 4.25 & -0.09 & -0.41 \\
LSTM-RGCN \cite{li2021modeling} & 49.46 & 1.10 & 1.07 & 6.30 & 17.73 & 1.20 & 49.30 & 1.59 & 0.95 & 15.36 & 14.56 & 1.54 & 49.02 & 0.90 & 1.42 & 4.93 & 5.24 & 0.11 \\
SFM \cite{zhang_stock_2017} & 49.24 & 2.81 & 2.51 & 18.37 & 2.41 & 2.29 & 49.30 & 3.37 & 3.47 & 19.85 & 21.17 & 2.10 & 49.01 & 1.40 & 1.51 & 7.38 & 6.71 & 0.95 \\
HATR-I \cite{wang2022hatr} & 49.38 & 0.41 & 0.68 & 1.90 & 4.43 & 0.01 & / & / & / & / & / & / & 49.08 & 3.89 & 2.67 & 16.81 & 21.34 & 1.64 \\
THGNN \cite{xiang2022temporal} & 50.75 & 0.76 & 0.63 & 3.73 & 9.78 & 0.36 & 50.79 & 4.20 & 4.25 & 15.36 & 21.83 & 1.81 & 49.59 & 0.67 & 0.72 & 2.76 & 4.97 & 0.08 \\
SAMBA \cite{samba} & 50.68 & 1.04 & 0.84 & 5.59 & 13.13 & 0.73 & 50.58 & 2.55 & 1.50 & 13.84 & 20.09 & 1.66 & 50.30 & 0.70 & 0.59 & 2.55 & 11.43 & 0.50 \\
Ours w/o expert signals & 57.66 & 7.83 & 9.49 & 35.60 & 32.55 & 2.64 & 54.94 & 7.15 & 7.80 & 36.68 & 26.93 & 2.61 & 53.82 & 6.29 & 6.71 & 26.04 & 20.41 & 1.26 \\
Ours w/ expert signals & \textbf{57.83} & \textbf{7.98} & \textbf{9.69} & \textbf{36.66} & \textbf{32.91} & \textbf{2.69} & \textbf{55.02} & \textbf{8.26} & \textbf{8.29} & \textbf{43.17} & \textbf{29.56} & \textbf{2.92} & \textbf{54.81} & \textbf{7.62} & \textbf{7.78} & \textbf{30.69} & \textbf{28.46} & \textbf{1.87} \\
\hline
\end{tabular}
}
\end{table*}

\begin{table*}[htbp]
\centering
\caption{Performance comparison of the GAT model using different graph structures. ACC, IC, RIC, ICIR, AR are measured in percentage (\%).}
\label{tab:gat_graph_comparison}
\resizebox{\textwidth}{!}{
\begin{tabular}{c|cccccc|cccccc|cccccc}
\hline
& \multicolumn{6}{c|}{NASDAQ 100} & \multicolumn{6}{c|}{SP 500} & \multicolumn{6}{c}{Updated StockNet} \\
\cline{2-7} \cline{8-13} \cline{14-19}
Graph & ACC & IC & RIC & ICIR & AR & SR & ACC & IC & RIC & ICIR & AR & SR & ACC & IC & RIC & ICIR & AR & SR \\
\hline
Correlation & 57.15 & 9.09 & 7.02 & 24.51 & 27.73 & 1.66 & 53.08 & 6.43 & 7.53 & 30.46 & 24.95 & 2.04 & 53.61 & 5.93 & 5.87 & 20.76 & 23.77 & 1.56 \\
Industry & 54.44 & 4.42 & 3.29 & 16.93 & 16.94 & 0.76 & 52.61 & 5.53 & 5.15 & 41.14 & 19.15 & 1.56 & 52.67 & 3.91 & 5.19 & 19.91 & 15.65 & 0.89 \\
Industry+Correlation & \textbf{57.83} & \textbf{7.98} & \textbf{9.69} & \textbf{36.66} & \textbf{32.91} & \textbf{2.69} & \textbf{55.02} & \textbf{8.26} & \textbf{8.29} & \textbf{43.17} & \textbf{29.56} & \textbf{2.92} & \textbf{54.81} & \textbf{7.62} & \textbf{7.78} & \textbf{30.69} & \textbf{28.46} & \textbf{1.87} \\
\hline
\end{tabular}
}
\end{table*}

\begin{table*}[htbp]
\centering
\caption{Performance comparison between GCN and GAT. ACC, IC, RIC, ICIR, AR are measured in percentage (\%).}
\label{tab:model_type_comparison}
\resizebox{\textwidth}{!}{
\begin{tabular}{c|cccccc|cccccc|cccccc}
\hline
& \multicolumn{6}{c|}{NASDAQ 100} & \multicolumn{6}{c|}{SP 500} & \multicolumn{6}{c}{Updated StockNet} \\
\cline{2-7} \cline{8-13} \cline{14-19}
Model type & ACC & IC & RIC & ICIR & AR & SR & ACC & IC & RIC & ICIR & AR & SR & ACC & IC & RIC & ICIR & AR & SR \\
\hline
GCN & 56.99 & 5.60 & 7.09 & 27.66 & 28.12 & 2.31 & 54.83 & 7.87 & 5.43 & 41.51 & 23.52 & 2.19 & 54.06 & 5.85 & 4.56 & \textbf{34.05} & 20.27 & 1.29 \\
GAT & \textbf{57.83} & \textbf{7.98} & \textbf{9.69} & \textbf{36.66} & \textbf{32.91} & \textbf{2.69} & \textbf{55.02} & \textbf{8.26} & \textbf{8.29} & \textbf{43.17} & \textbf{29.56} & \textbf{2.92} & \textbf{54.81} & \textbf{7.62} & \textbf{7.78} & 30.69 & \textbf{28.46} & \textbf{1.87} \\
\hline
\end{tabular}
}
\end{table*}

\subsection{Ablation Study}

In this ablation study, we analyze the impact of different components on the performance of our model. 

For the graph structure, we tested three configurations using the GAT architecture: employing only the industry graph, only the correlation graph, and a combination of both by our DualGAT. The results shown in Table \ref{tab:gat_graph_comparison} indicated that the combined graph setup outperformed the individual setups, demonstrating the effectiveness of integrating diverse market relationships into the model. Meanwhile, for a single graph source, we find that combining the information contained in the pre-trained model output with the expert signals yields better results than the message passing using only expert signals, especially in terms of AR and SR. This further demonstrates that expert signals can work synergistically with traditional financial features.

We also compared the performance of two graph neural network architectures, GCN and GAT, using the combined industry and correlation graphs. As shown in Table \ref{tab:model_type_comparison} ,the GAT architecture showed superior performance over GCN.

\textcolor{black}{Beyond the architectural components, we also evaluated the impact of expert signal quality on our framework. We fed our DualGAT model with expert signals generated by various baseline expert identification algorithms shown in Figure \ref{fig:pre_experiment1}. Our proposed algorithm consistently outperforms all baseline methods, confirming its critical contribution to the model's superior performance. Detailed results are available in Supplementary Material S9.}

\section{Discussion}
Our study demonstrates a significant advancement in utilizing social media data for stock prediction by effectively addressing two fundamental challenges: identifying valuable expert signals from noisy social media and propagating these sparse signals across the stock network for better prediction.

A critical aspect of our framework is its robustness in handling sparse expert signals across different sectors and stock types. Our dual-graph architecture effectively addresses the sparsity problem through different propagation mechanisms. We have found that the industry graph enables each stock's signal to reach an expected 20.65\% of nodes in a single hop while the correlation graph complements this by propagating signals to an additional 9.16\% of nodes. This dual-graph approach ensures comprehensive and sufficient signal coverage around 90\% after two-hop propagation. \textcolor{black}{Empirical analysis further confirms that two-hop propagation achieves optimal performance (see Supplementary Material S10).}
The accuracy of signal propagation is equally crucial as coverage. Our analysis reveals strong trend alignment between connected stocks, supporting the reliability of our propagation mechanism. In the correlation graph, stock pairs meeting our correlation thresholds (0.67 and 0.77) show trend alignment probabilities of 75.09\% and 80.04\% respectively in the next day. Similarly, stocks connected in the industry graph demonstrate a 70.4\% probability of aligned trends. Even accounting for potential error propagation, the accuracy of the propagated expert signals remains robust, with even the simplest message passing approach outperforming baseline models in Table \ref{tab:performance_evaluation_models}.

While our framework excels at daily predictions, it's important to know its scope. Our focus is on daily prediction horizons because price and social media information are most relevant at this timescale. Factors that mainly influence long-term predictions, such as policy changes and macroeconomic conditions, fall outside our scope. Nevertheless, we can still leverage social media to more accurately capture and utilize the impact of these factors on predictions. For instance, when there is disagreement among users about the stock trend prediction due to policy changes, the effects of these factors are often reflected in social media discussions, leading to a surge in activity where experts may actively participate. Figure \ref{fig:pre_experiment1} shows that experts often provide more accurate insights than the general sentiment of the crowd. By identifying and leveraging these expert opinions, our framework captures the actionable impact of these events more effectively, ultimately improving prediction accuracy. On the other hand, the primary focus of our work is based on social media which could indirectly capture the effects of these external factors. Based on this, our method can work synergistically with other factors to improve overall prediction performance.

Ultimately, our method emphasizes the use of expert opinions based on self-labeled sentiments, which avoids the computational overhead associated with large language models (LLMs). While LLMs could be applied for sentiment analysis, our approach is demonstrably more computationally efficient and better suited for real-time applications. \textcolor{black}{In fact, our entire framework, from signal extraction to the DualGAT architecture, is designed for high efficiency. Empirical analysis shows that our DualGAT model's inference time exhibits a growth rate significantly below linear scaling with the number of stocks, ensuring its practicality for large-scale use. A detailed computational and scalability analysis is provided in Supplementary Material S11.} As such, our approach does not compete with LLMs but complements them by focusing on meaningful opinions. For instance, one promising direction involves using our expert tracing system to quickly identify potential experts in daily tweets. This enables prioritizing their texts for LLM processing, allowing LLMs to extract information beyond sentiment. This selective processing enhances synergy between LLM capabilities and expert signals.

\section{Conclusion}
Our research has highlighted the significant noise present in public sentiment signals derived from social media platforms like StockTwits, which complicates the forecasting of stock prices. To tackle this challenge, we developed an expert tracing system that effectively identifies and utilizes valuable expert signals amidst this noise. While these signals are highly effective, their sparsity limits their direct application. To overcome the limitations posed by sparse signals, we constructed a dual graph-based framework that employs graph attention networks. This innovative approach allows for the effective propagation of sparse expert signals across the network, significantly enhancing the richness and accuracy of our predictions. Our method significantly outperforms various benchmark models in time-series forecasting and stock prediction.

For future work, there is substantial potential to further exploit the wealth of information in social media content. As StockTwits dataset contains follower counts for each user, we could investigate the relationship between social media influence and stock price movements, and examine the predictive capabilities of influential users. Another promising direction involves processing textual content of expert or influencer tweets through large language models to generate embeddings. These embeddings could be integrated with sentiment features, providing richer features for our graph-based models.


\bibliographystyle{IEEEtran}
\bibliography{Reference}

\end{document}